\documentclass{pnastwo}

\usepackage[normalem]{ulem}
\usepackage{amsmath, amsfonts, amssymb}
\usepackage{color}
\usepackage[latin1]{inputenc}       
\usepackage[T1]{fontenc}            
\usepackage{float}

\usepackage[modulo]{lineno}

\usepackage{graphicx}        
\usepackage{listings}
\contributor{Submitted to Proceedings
of the National Academy of Sciences of the United States of America}
\url{}
\copyrightyear{2008}
\issuedate{Issue Date}
\volume{Volume}
\issuenumber{Issue Number}

\begin{document}

\title{Does ``model-free'' forecasting really outperform the ``true'' model?} 

\author{Florian Hartig\affil{1}{University of Freiburg, Biometry and Environmental System Analysis, Freiburg, Germany}, Carsten Dormann\affil{1}{}}

\contributor{Submitted to Proceedings of the National Academy of Sciences
of the United States of America}

\maketitle

\begin{article}

\linenumbers

Estimating population models from uncertain observations is an important problem in ecology. Perretti et al. observed that standard Bayesian state-space solutions to this problem may provide biased parameter estimates when the underlying dynamics are chaotic \cite{Perretti-Modelfreeforecasting-2013}. Consequently, forecasts based on these estimates showed poor predictive accuracy compared to simple ``model-free'' methods, which lead Perretti et al. to conclude that ``Model-free forecasting outperforms the correct mechanistic model for simulated and experimental data''. However, a simple modification of the statistical methods also suffices to remove the bias and reverse their results.

\begin{figure}
\centering
\includegraphics [width=6.5cm]{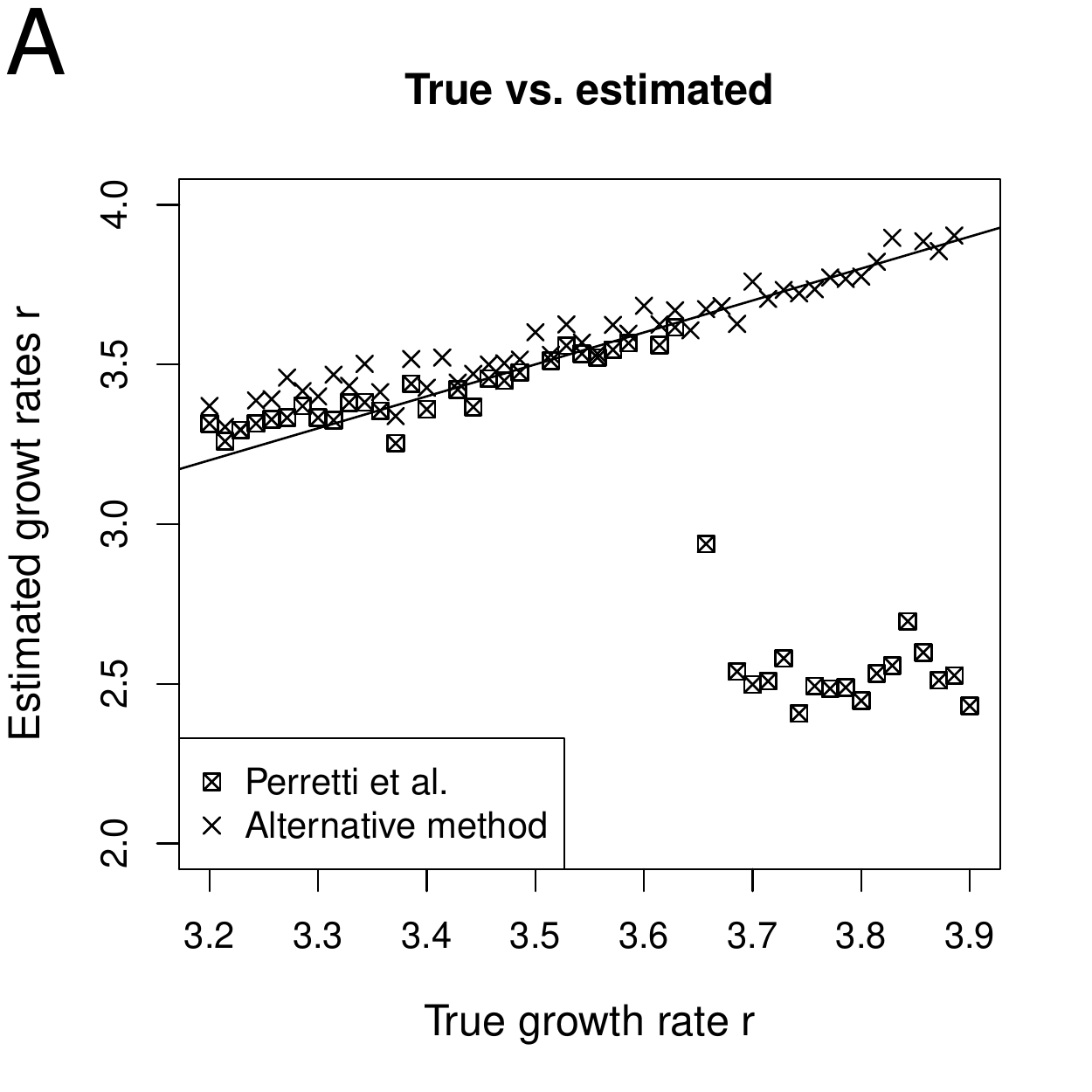} 
\includegraphics [width=6.5cm]{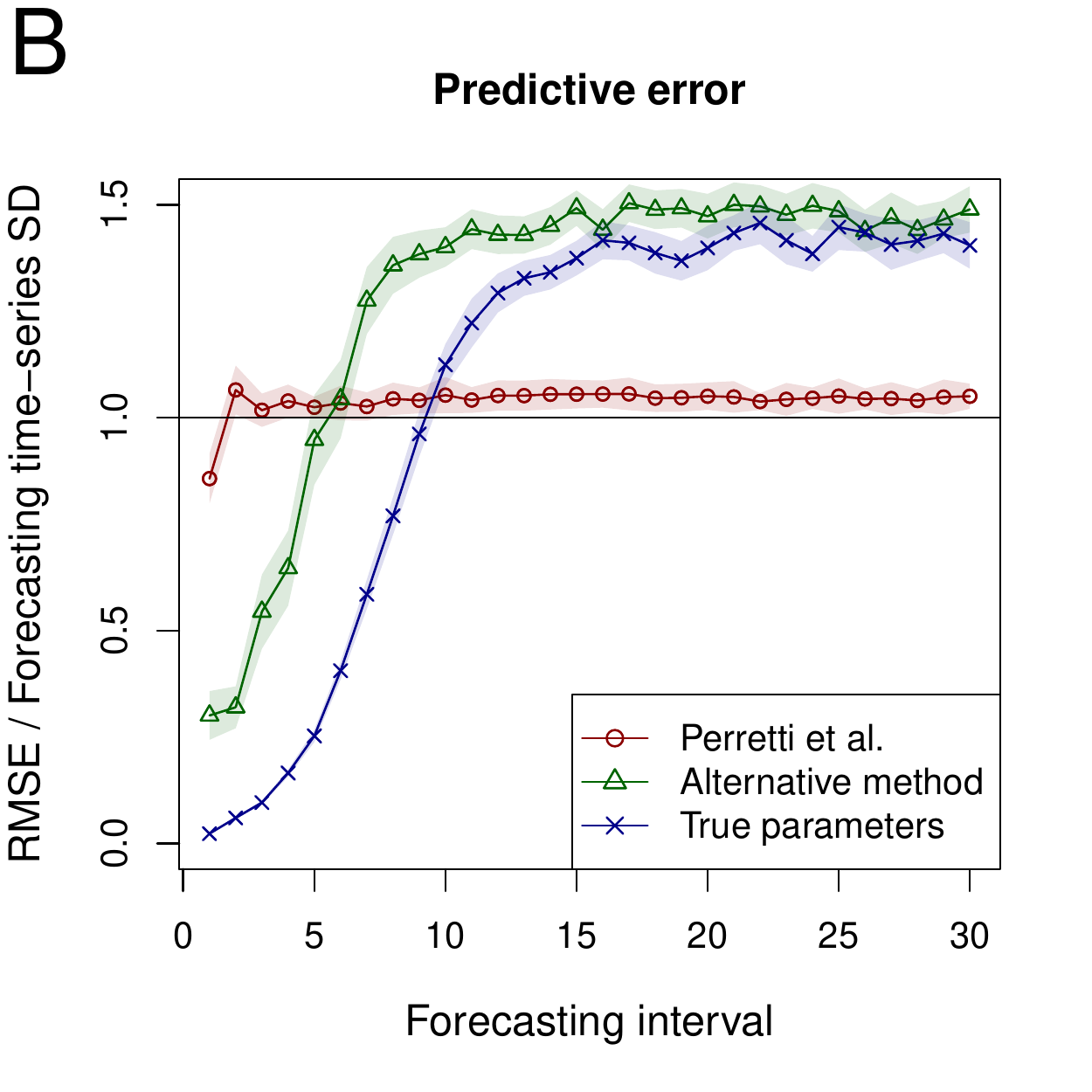} 
\caption{Comparison of median posterior parameter estimates for different values of the groth rate $r$ (A) and predictive error for $r=3.7$ (B) for the estimation method used by Perretti et al. and our alternative estimation method. For the lower panel we show additionally the predictive error of the true parameters. Predictive error is measured by standardized RMSE \cite{Perretti-Modelfreeforecasting-2013}, shaded area depicts the 95\% CI. Note that biased, but stable $r$-values estimated by Perretti et al. have larger short-term, but smaller long-term error than the true values. \label{figure: modelPredictions}} 
\end{figure}

The instability of both maximum likelihood and Bayesian inference for chaotic models has been recognized before \cite{Pisarenko-Statisticalmethodsparameter-2004}. Deterministic chaos produces quasi-random trajectories that are extremely sensitive to changes in parameters and initial conditions. Likelihoods are therefore often highly irregular in shape \cite{Wood-Statisticalinferencenoisy-2010}. Moreover, if there is sufficient noise in either process or observations, ``true'' chaotic parameters may have lower likelihood than alternative parameters with stable trajectories, effectively devaluating maximum likelihood as a consistent estimator for chaotic dynamical systems \cite{Judd-Failuremaximumlikelihood-2007}. The reason is that for chaotic models, the smallest amount of noise leads to diverging population trajectories, so that simulations from the same parameters may be further apart from each other on the long run than from a stable trajectory at the time-series mean (see Fig.~\ref{figure: modelPredictions}B).

There are a number of known methods to bypass these problems. Using summary statistics, potentially in an approximate Bayesian framework, is one of them \cite{Wood-Statisticalinferencenoisy-2010, Hartig-Statisticalinferencestochastic-2011}. In the case of the chaotic models presented by Perretti et al., however, there is a simpler solution. Because the bias arises from the long-term divergence of the chaotic population dynamics, a simple solution is to divide the time series in smaller subsets, and fit the model to those individually \cite{Pisarenko-Statisticalmethodsparameter-2004}. We applied this method to the example of the logistic model used in Perretti et al. and obtained parameter estimates that are virtually unbiased. The resulting median parameter estimates show similar dynamics and predictive uncertainty as the "true" model, with lower short-term error than the statistical alternatives (Fig.~\ref{figure: modelPredictions}, further details see supplementary material). It seems likely to us that similar results could also be obtained for the other model types examined in \cite{Perretti-Modelfreeforecasting-2013}. 


Perretti et al. highlight a statistical problem in inferring parameters of chaotic dynamics, which is very important, as it is conceivable that the described bias may have gone unnoticed when working with empirical data only. However, our simulations question Perretti et al.'s conclusion that these problems fundamentally render ``model-free'' approaches superior. As we show, using a simple modification of the statistical method provides a better solution to the problem, without having to give up other advantages of process-based models that might also benefit forecasting in the long run, such as transferability and theoretical understanding.

\end{article}

\setcounter{figure}{0}
\makeatletter 
\renewcommand{\thefigure}{S\@arabic\c@figure}

\newpage

\linenumbers

\section{Supplementary material}

Models can be used for various purposes. The most important purpose in science is arguably to summarize and test our understanding of a system. Another important purpose, however, is to make predictions. It is often implicitly assumed that those two purposes are aligned, in the sense that a "more correct" model should also make better predictions. Perretti et al. rightly challenge this intuition \cite{Perretti-Modelfreeforecasting-2013}. They simulate data from chaotic population models, which ensures that we know the data-creating mechanism. They then compare both statistical time series models that do not impose any particular model structure and the estimated true (structurally correct) models regarding their predictive error. Their results seem to show that the predictive accuracy of the true models, with parameters fit in a Bayesian state-space framework, is worse than that of simple time-series models. They conclude that "Considering the recent push for ecosystem-based management and the increasing call for ecological predictions, our results suggest that a flexible model-free approach may be the most promising way forward." While there generally may well be a trade-off between structural correctness and predictive power, at least the logistic map, one of the examples chosen, does not support this claim. In the following, we discuss in more detail the problems and possible solutions for fitting the logistic map, the simplest in a set of four population models examined by \cite{Perretti-Modelfreeforecasting-2013}, and show that an alternative statistical estimation methods is able to deal with all problems that seemingly suggest the use of "model-free" methods over the true, mechanistic model. We believe that the reason for the low predictive accuracy is essentially the same for all models, namely the problem in the likelihood due to the chaotic dynamics mentioned in the main text. Our conclusions as well as our solution for this problem should therefore transfer to the other model types as well.

\subsection{Model specification}

The logistic model is used by \cite{Perretti-Modelfreeforecasting-2013} in its time-discrete version (often also called the logistic map), to which a lognormal process error is multiplied.

\begin{equation}\label{eq: population model}
N_{t+1} \sim \left(N_t \cdot r \cdot (1-N_t/K)\right) \cdot \gamma(0,\sigma^{proc})
\end{equation}

\begin{figure*}[hbt]
\centering
\includegraphics [width=7cm]{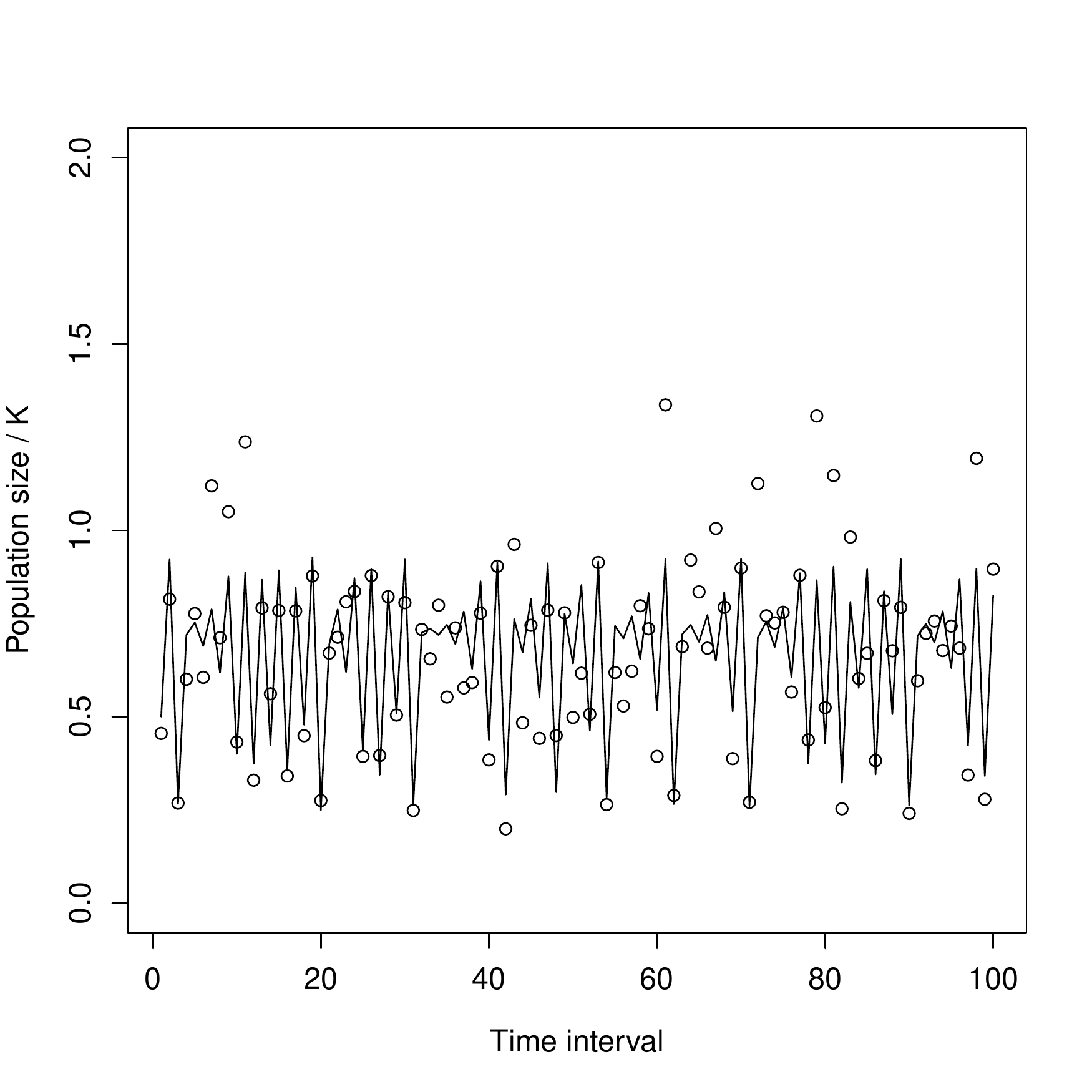} 
\caption{Population dynamics for $r=3.7$, $K=1$, $\sigma^{proc} =0.005$ and $\sigma^{obs} = 0.2$. Solid line shows the true population size, circles show observations. \label{figure: populationdynamics}} 
\end{figure*}

Here, $N_t$ is the population size at time $t$, $r$ is the intrinsic growth rate, $K$ is the carrying capacity, and $\gamma$ is a lognormal random variable with the standard parametrization of $\mu$ being the mean on the log-scale. We write $\sim$ to denote the stochastic process on the right side of the equation. It is assumed that the true population size $N$ is observed with error, leading to an observed population size $N_t^{obs}$ at time $t$ of

\begin{equation}\label{eq: observation model}
N_t^{obs} \sim \gamma(\mu = log(N_t),\sigma = \sigma^{obs})
\end{equation}

where $\gamma$ is again a lognormal random variable. Note that the lognormal distribution is skewed, and that the median of the $\gamma$ on the normal scale is $exp(\mu)$, but the expectation value is $E(X) = exp(\mu + 1/2 \sigma^2)$, meaning that both observation and process error create noise that is as likely to lead to higher than to lower values compared to the deterministic model, but with an expectation value that tends towards higher values. An example of the resulting population dynamics for the standard parameterization used in \cite{Perretti-Modelfreeforecasting-2013} is shown in Fig.~\ref{figure: populationdynamics}.

\subsection{JAGS model specification}

Similar to \cite{Perretti-Modelfreeforecasting-2013}, we created observations of a population time series with 100 time intervals from the population model eqs.~\ref{eq: population model},\ref{eq: observation model}. The first 50 data points were used to fit the model, and the second 50 data points to evaluate the model predictions. We fit the parameters with JAGS. Our JAGS model specification replicating Perretti et al. is displayed in Code~\ref{Source: originalJagsModified}. We refer to results created by this code as ``Perretti et al.''.

\begin{lstlisting}[language=R,caption=JAGS model specification for the model structure of Perretti et al., label=Source: originalJagsModified]
model{
    
    r ~ dunif(2,4.5)
    K ~ dunif(0.01,10)
  
    N0 ~ dunif(0.2,1.5)
    pop[1] <- N0
    
    ProcessPrecision <- 1/0.005/0.005
    ObservationPrecision ~ dunif(0,40)
    obsSD <- sqrt( 1 / ObservationPrecision)
    
    observed[1] ~ dlnorm(log(pop[1]), ObservationPrecision)
    
    for(Time in 1:(nObs-1))
    {   

      logpop[Time+1] <- log(max(0, r * pop[Time] * (1 - pop[Time] / K) ))

      pop[Time+1] ~ dlnorm(logpop[Time+1], ProcessPrecision) 
      
      observed[Time+1] ~ dlnorm(log(pop[Time+1]), ObservationPrecision)
      
    }
  }
  
\end{lstlisting}

As mentioned in the main text, it was noted before that this model specification, or to be more exact, the likelihood for this model, may fail to provide correct parameter estimates \cite{Pisarenko-Statisticalmethodsparameter-2004, Judd-Failuremaximumlikelihood-2007} due to a particularity in chaotic population dynamics. The reason is that, if starting conditions are not perfectly known, or if there is a process-error, population trajectories of identical parameters diverge to very different values after few time steps. Essentially, this is the definition of deterministic chaos. 

In consequence, even for the correct parameters, the expected long-term difference between predicted and observed values is higher than for a model that provides stable population dynamics at the time series mean (see Fig.~\ref{figure: modelPredictions}, main text). To explain this mathematically, imagine the chaotic time series shows a normal distribution of population sizes with standard deviation $\sigma$. The long-term predictive errors of a model defined by the time series mean will thus have a standard deviation of $\sigma$. However, the long-term predictive errors of the true model, which has the same variance as the data, but is uncorrelated to the time series after a while, will have a standard deviation of $\sqrt{2} \sigma$ according to standard arguments for adding random variables, which is exactly what we see in Fig.~\ref{figure: modelPredictions}. Thus, deterministic chaos leads to the paradoxical situation that the ``true'' model is worse at making long-term predictions than the statistical mean if one looks at expected difference between observed and predicted population sizes only and neglects that the mean as a model obviously misrepresents other aspects of the time-series such as the variability or differential changes.

Pisarenko and Sornette examined the problems of parameter estimation bias that result from this phenomenon and concluded that a simple solution is to "break" the time series data in smaller intervals, which are estimated individually \cite{Pisarenko-Statisticalmethodsparameter-2004}. Strictly speaking, the resulting model is not the "true" likelihood for the problem, and there is certainly a great deal to be said about the philosophical and statistical justification for this approach, but on the practical side \cite{Pisarenko-Statisticalmethodsparameter-2004} showed that this approach avoids the instability in the likelihood that originates from the divergence of the chaotic population dynamics on the long run and generally leads to good parameter estimates. Code~\ref{Source: modifiedJags} shows the JAGS model specification using this approach. The model is identical, except for the fact that the observed data, originally a 50 step time series, was split up in a 5x10 matrix, and each 5-step interval is fitted individually, assuming the initial value is unknown.  We will refer to this code as the ``alternative model specification''.    

\begin{lstlisting}[language=R, caption=JAGS model specification of the modified model, label=Source: modifiedJags]
model{
  
    # Priors
    
    r ~ dunif(2,4.5)
    K ~ dunif(0.01,10)
    
    ProcessPrecision <- 1/0.005/0.005
    ObservationPrecision ~ dunif(0.1,40)
    obsSD <- sqrt( 1 / ObservationPrecision)

    # Likelihood
    
    for(i in 1:steps)
    {
      pop[1,i] ~ dunif(0.2,1.5)
      observed[1,i] ~ dlnorm(log(pop[1,i]), ObservationPrecision)
      
      for(j in 1:(substeps-1))
      {   
        procerr[j,i] ~ dlnorm(0, ProcessPrecision)
        
        pop[j+1,i] <- max(0, r * pop[j,i] * (1 - pop[j,i] / K) ) * procerr[j,i]
        
        observed[j+1,i] ~ dlnorm(log(pop[1+j,i]), ObservationPrecision)
      }
    }
  }
  
\end{lstlisting}

For all results, we ran three chains and checked convergence by visual inspection of the chain and Gelman-Rubin plots, as well as by the Gelman-Rubin multivariate potenial scale reduction factor (mpsrf), which we required to be $<1.2$. For the model specification used in Perretti et al., chain mixing remained poor, and convergence therefore extremely slow. In principle, it is not uncommon that likelihood functions for chaotic systems exhibit extremely irregular shapes that are a challenge for MCMC samplers \cite{Wood-Statisticalinferencenoisy-2010}, however, in this case it was probably a problem of the sampler, because posterior samples were concentrated in unproblematic stable parameter regions. To compensate for the slow convergence, we ran 500.000 adaptation steps, 500.000 burn-in and 5.000.000 iterations for the posterior samples, which provided sufficient convergence. Chains were initialized with different, dispersed r-values to better test for convergence. For the alternative JAGS model, convergence was fast, so we generally ran three MCMC chains with 20.000 steps adaptation, 100.000 steps burn-in and 300.000 analysis steps.

\subsection{Results: Bias and predictive error for fixed r}

To analyze bias and predictive error for the chaotic parameter set of $r=3.7$, $K=1$, $\sigma^{proc} =0.005$ and $\sigma^{obs} = 0.2$ that was used in \cite{Perretti-Modelfreeforecasting-2013}, we created 50 test data sets with 100 time steps each from these parameters to reproduce the results for the logistic model, particularly Fig.~1, as well as Fig.~S1 in the supporting information. Parameters were fit for each scenario for the first 50 data points with the three different model specifications explained before. Additionally, we added a third option, the predictive error of the true parameters that were used to create the data. Basically, this option represents a benchmark by displaying how well an identical model would be at predicting "itself", and it allows to visualize the arguments made in the previous section about the ``true'' parameters showing larger long-term predictive errors than ``wrong'', but stable parameters. Here, as in the following experiment, we removed a small number of runs for which the multivariate potential scale reduction factor mpsrf was greated than 1.2, indicating that the MCMC had not probably not converged yet. 

\begin{figure*}
\centering
\includegraphics [width=5cm]{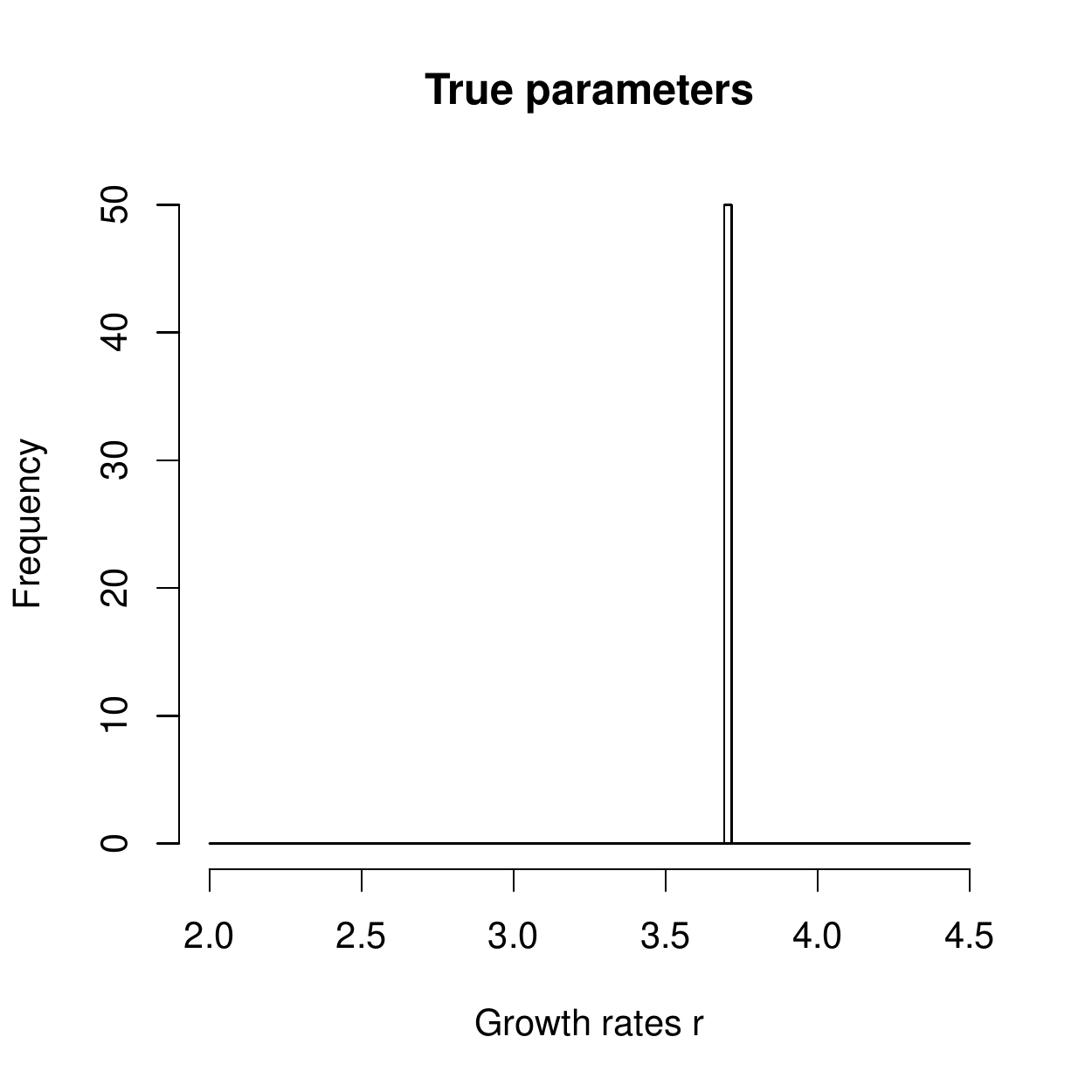}
\includegraphics [width=5cm]{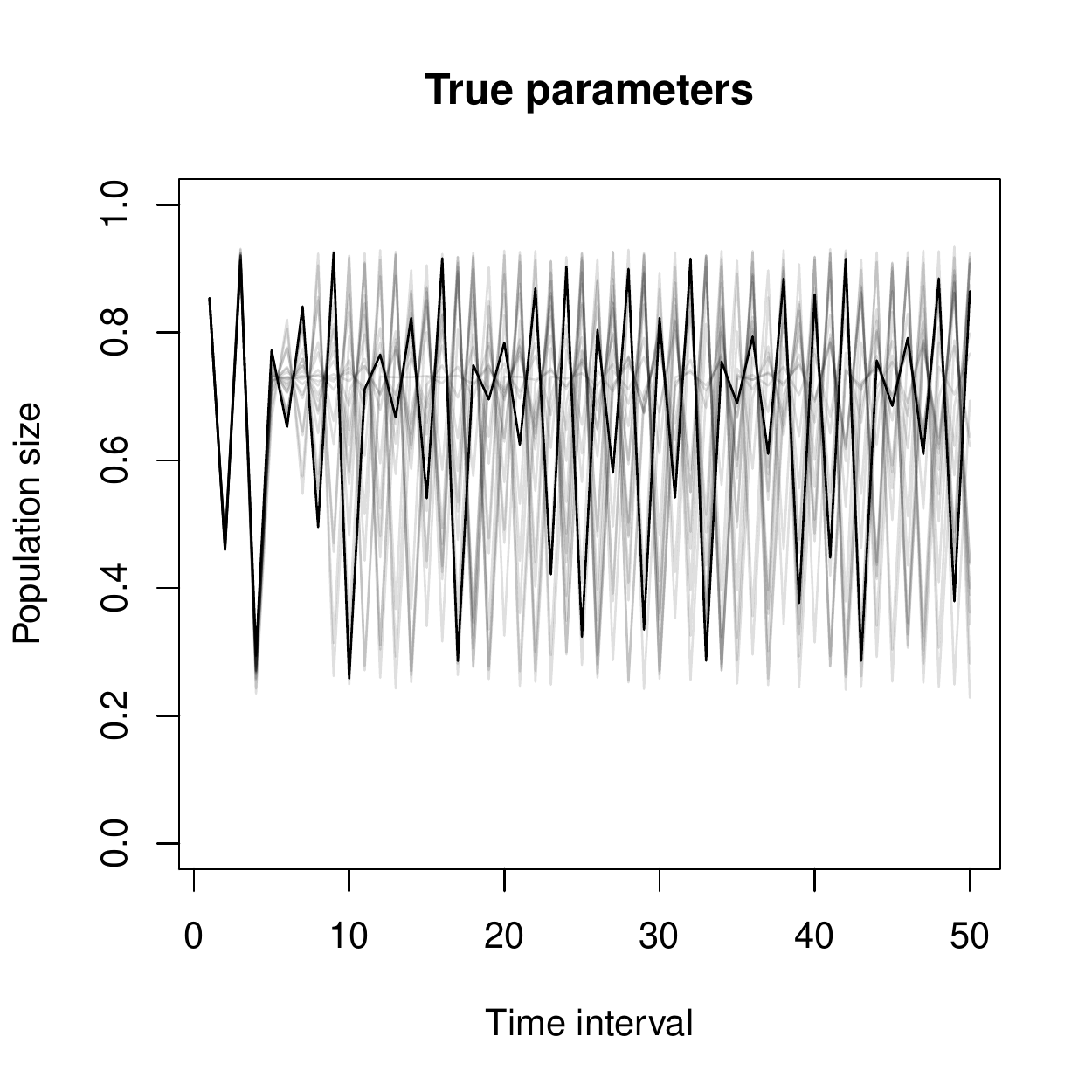} 
\includegraphics [width=5cm]{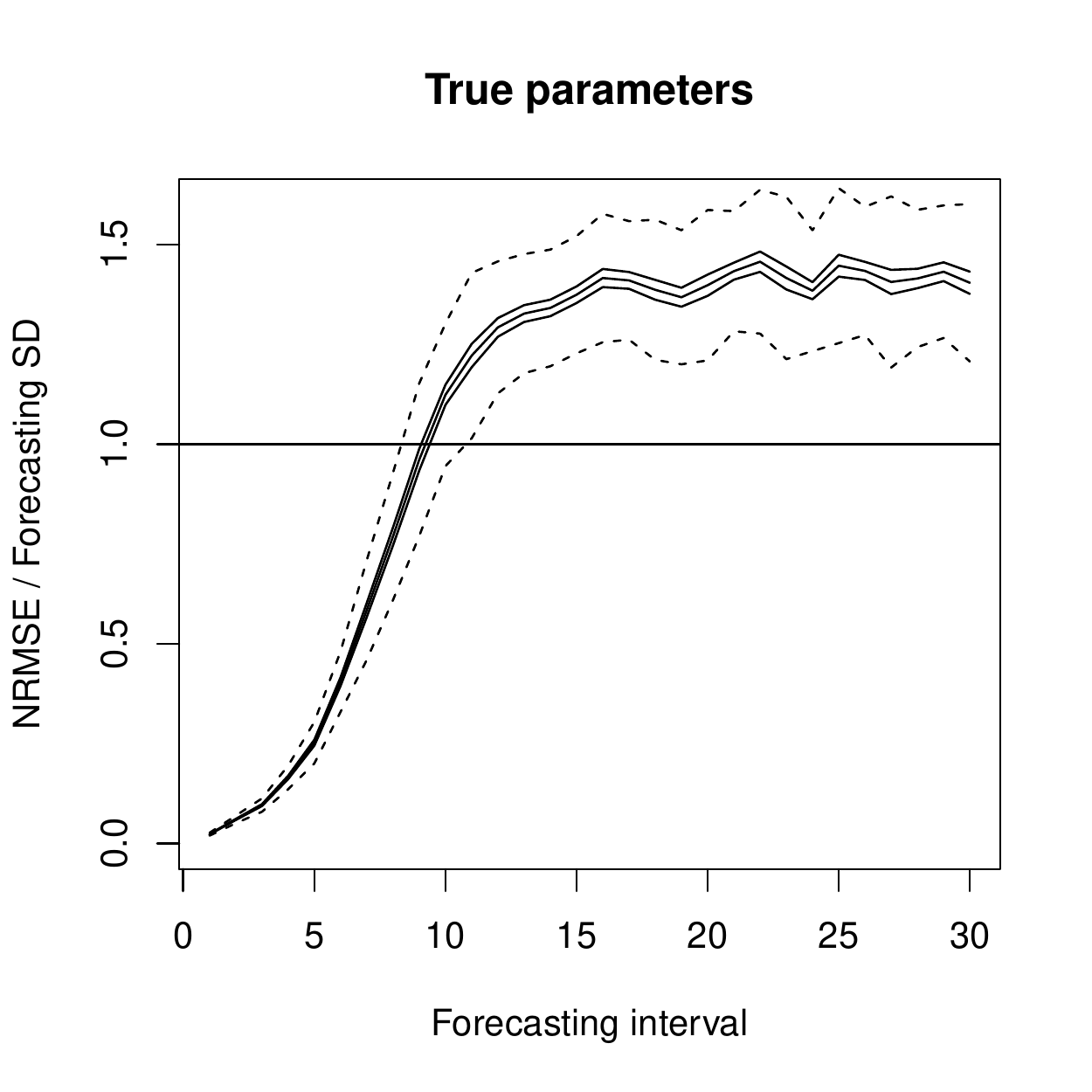} 
\includegraphics [width=5cm]{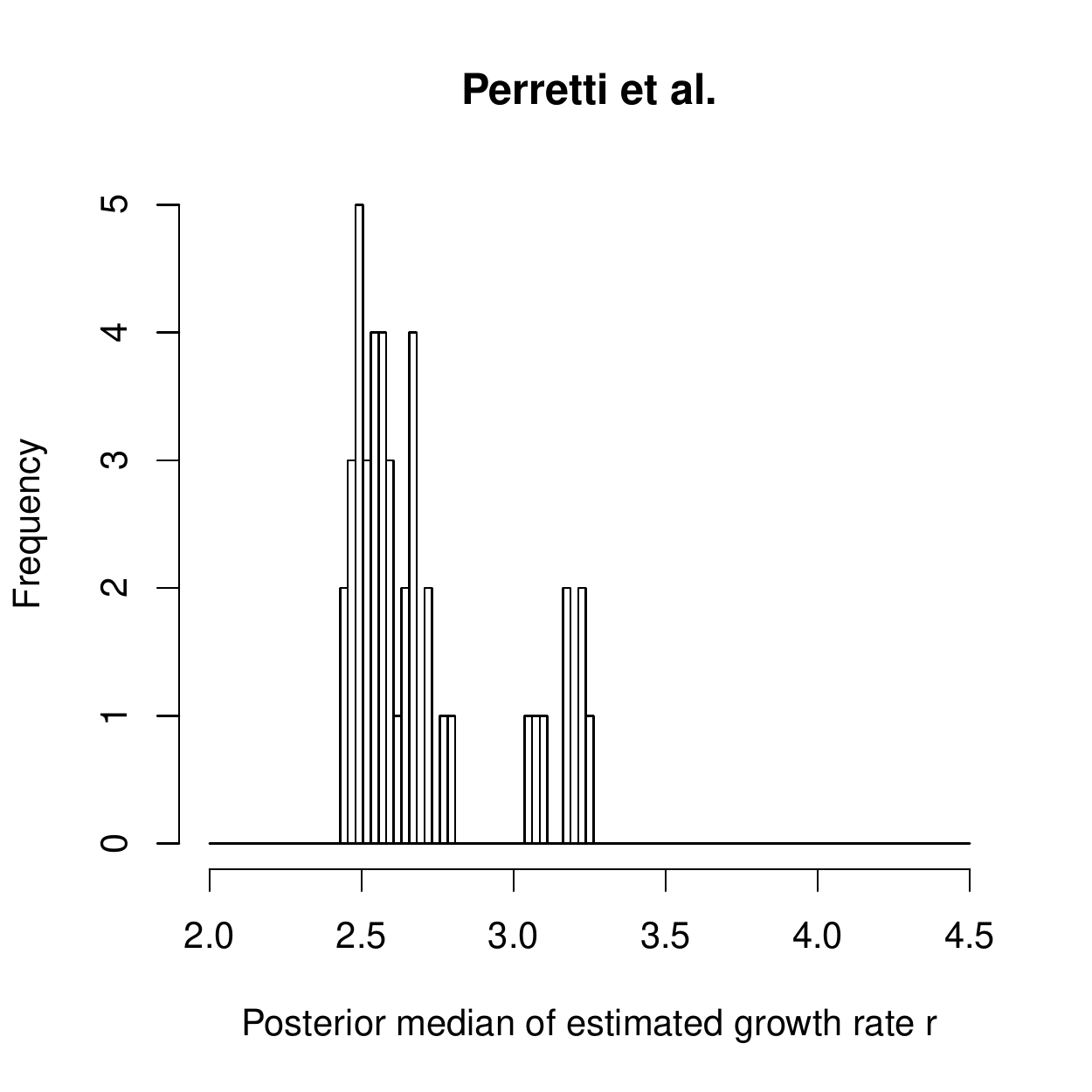}
\includegraphics [width=5cm]{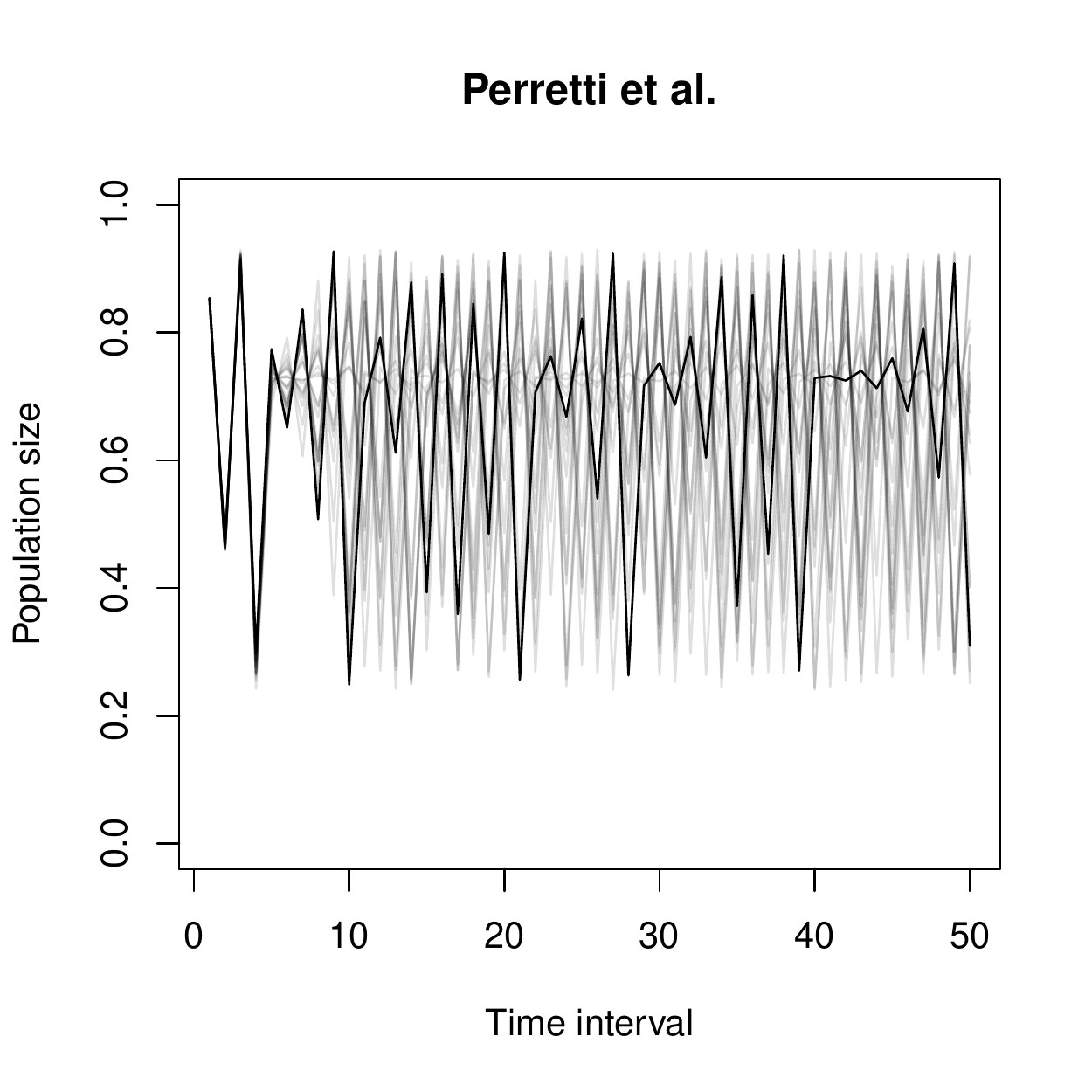} 
\includegraphics [width=5cm]{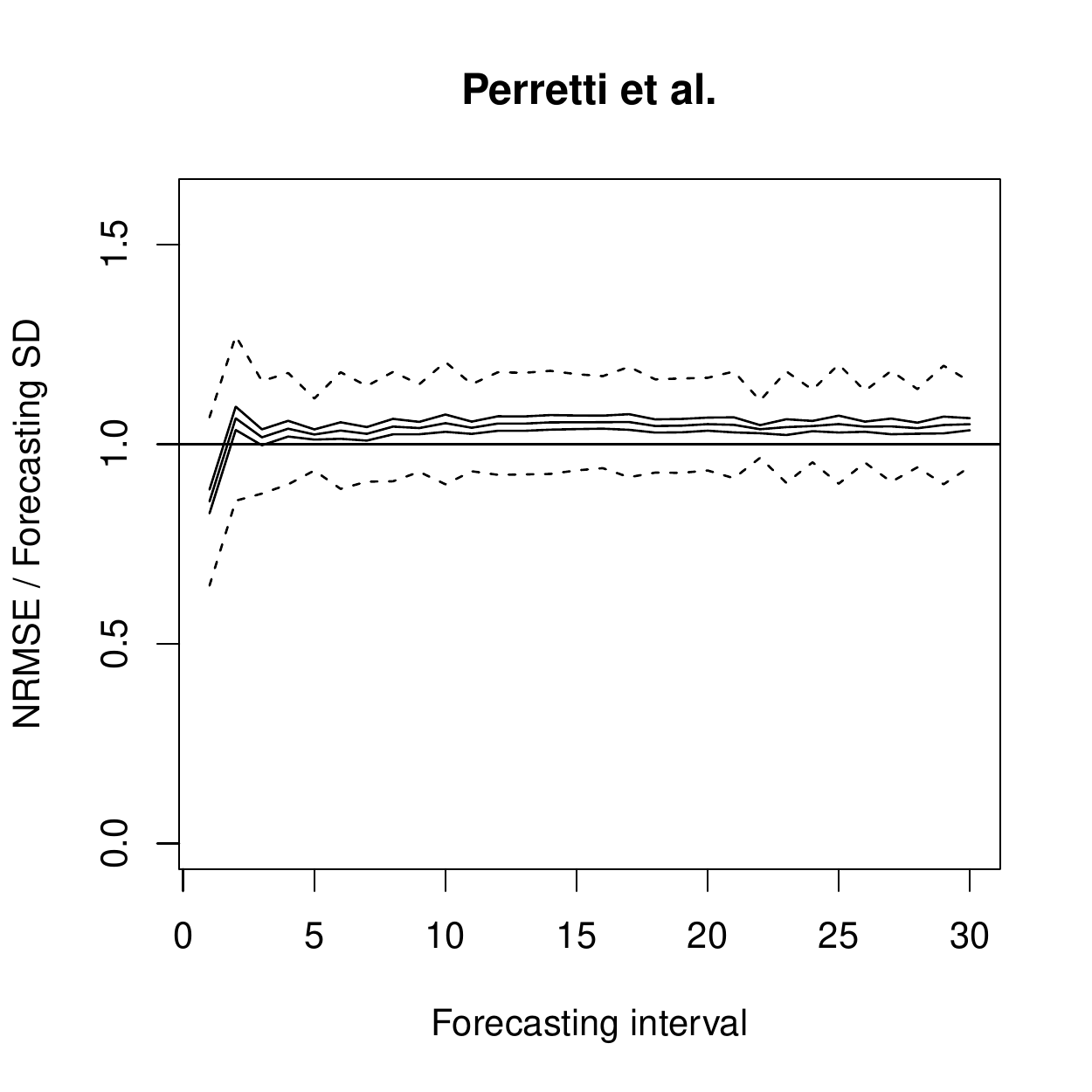} 
\includegraphics [width=5cm]{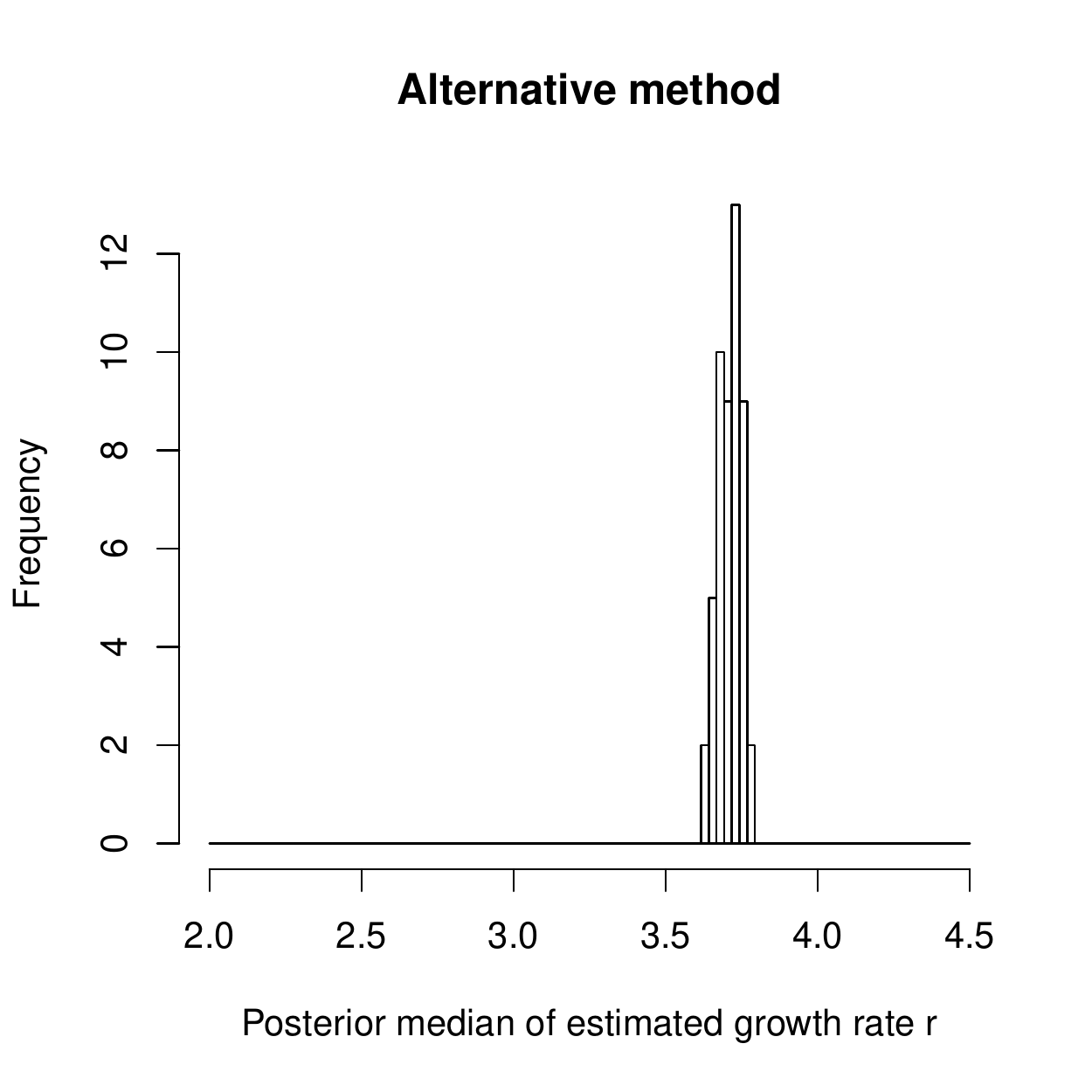} 
\includegraphics [width=5cm]{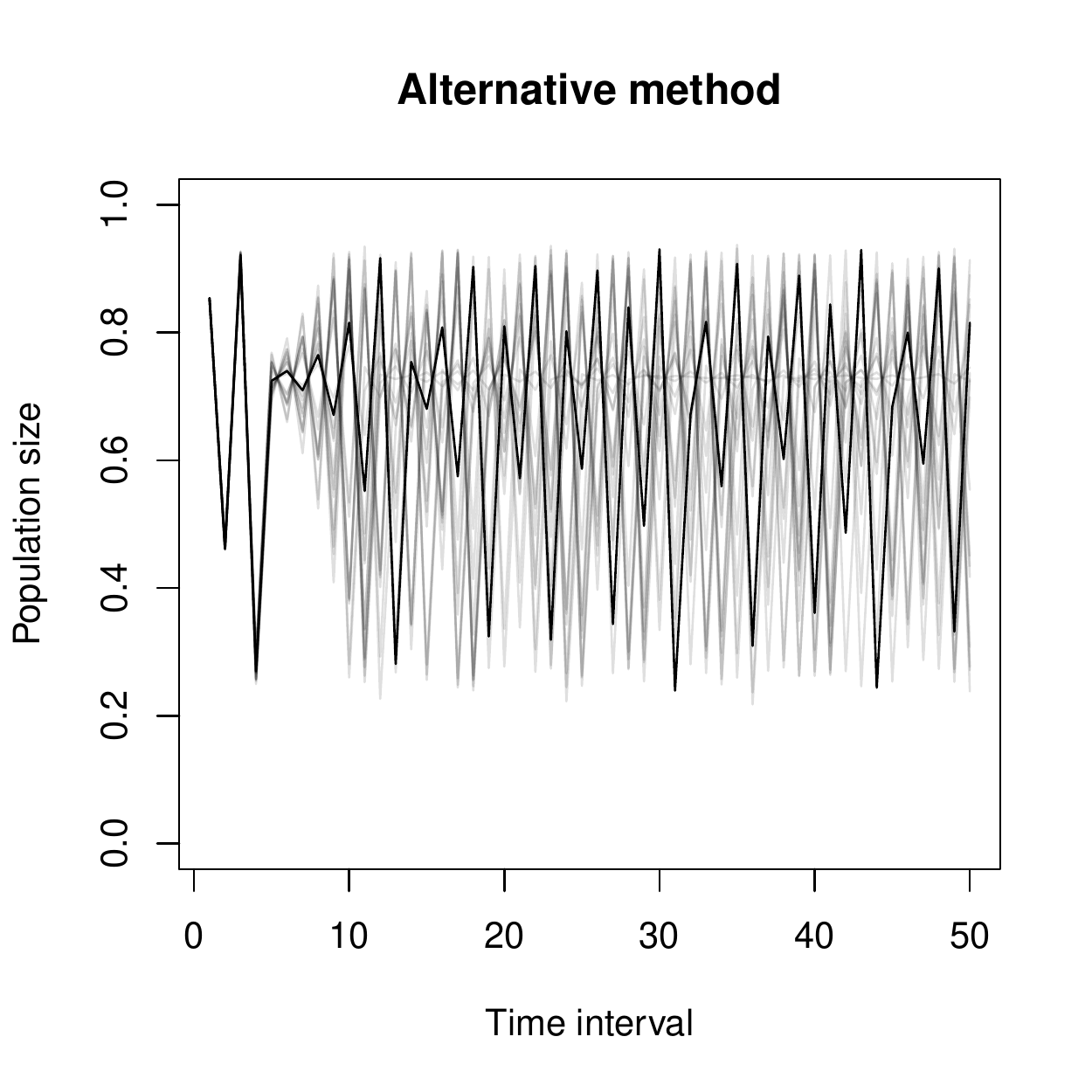} 
\includegraphics [width=5cm]{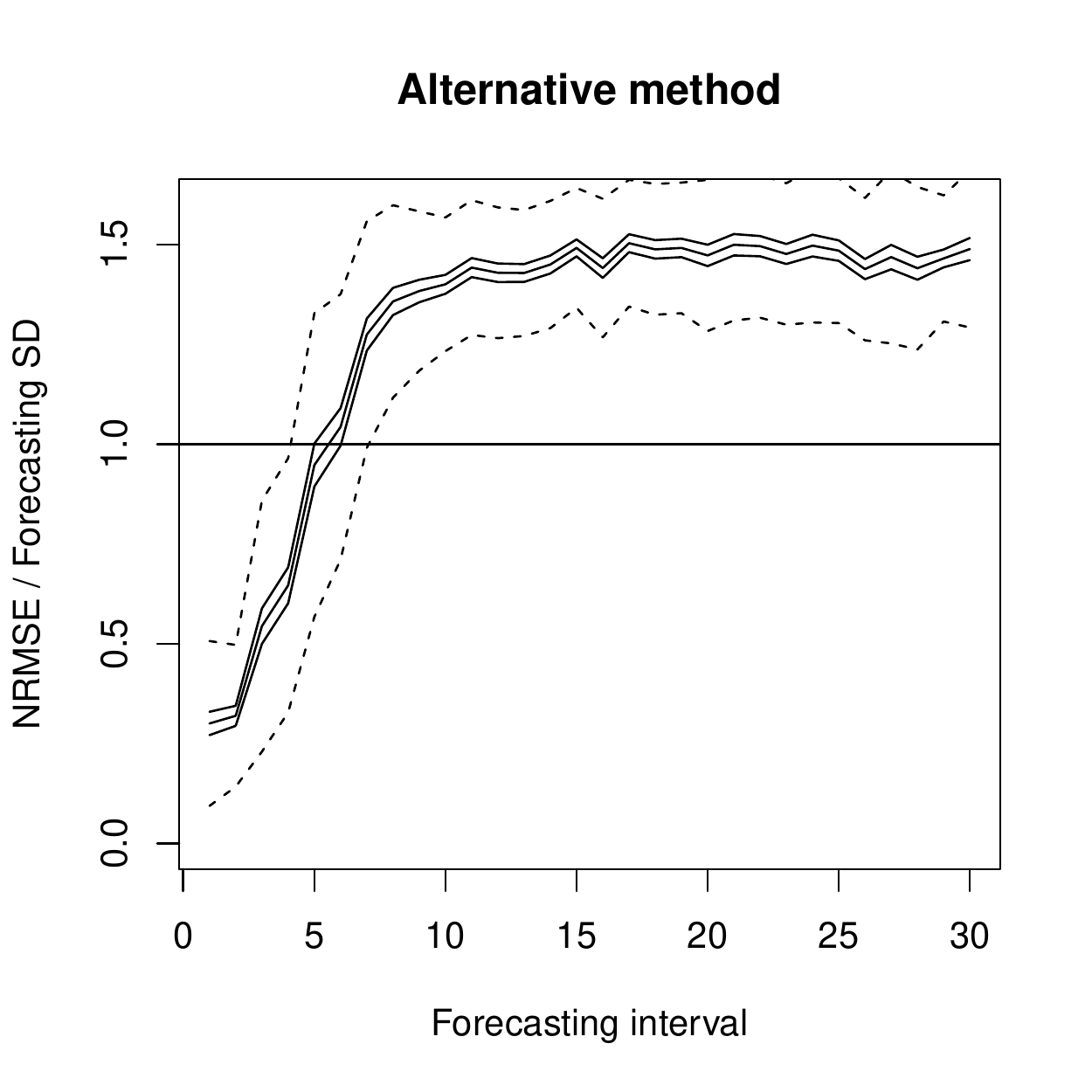}  
\caption{Histogram of median $r$ estimates (left column), population dynamics of the first 20 parameter estimates (gray line) compared to the true dynamics (black line) (middle column), and predictive error in terms of mean, standard error and standard deviation (right column) for true parameters (top), parameter estimation method of Perretti et al. (middle) and our alternative estimation method (bottom).\label{figure: predictive uncertainty}} 
\end{figure*}

We find that for the parameter estimation method of Perretti et al., posterior estimates are biased towards smaller values of $r$, for which population dynamics are stable. The resulting deficit of population variability as compared to the data is compensated for by an observation error that is estimated too high. This is the result one would expect from the findings of \cite{Pisarenko-Statisticalmethodsparameter-2004}. In contrast, our alternative parameter estimation method leads to clearly superior parameter estimates, both for $r$ (see Fig.~\ref{figure: predictive uncertainty}) and for the other parameters that are not shown here.

To derive predictions from the fitted models, we used median parameter estimates instead of the posterior mode that was used in the original paper, for two reasons. Firstly, the median is preferable because it is a more stable estimator than the mode for summarizing the MCMC output, which is, despite using indicators of convergence, a finite sample from the true posterior. And secondly, specially for the partly skewed and irregular distribution that we encountered here, we believe the median provides a better summary of the information in the posterior chain. However, we confirmed for all our results that similar results can be obtained using the mode. 

Predictive uncertainty was calculated as in the original paper (Perretti, personal communication): we first fit the model to the 50 calibration time steps of the 50 data sets. Then, for each data set $i$ and for each forecasting interval $p$, we predict the final population size for all possible $50-p$ intervals of length $p$ in the 50 validation time steps, using the true initial population size and the respective (posterior median or mode) parameters from the fit. This results in a set of predictive errors for the different time series, different forecasting intervals $p$ and the different intervals of length $p$ in the validation data. We calculate the root mean squared error (RMSE) across all errors from the same data set and the same forecasting interval, and if the forecasting interval is $p$, divide by the standard deviation of the validation time series, starting from the $50+p$ to the last time step, resulting in one value of the standardized root mean squared errors (SRMSE) for each data set and each forecasting interval. In the results, we show mean and and uncertainty over the 50 replicates of SRMSE for each forecasting interval.

The results show that, while the parameter estimation method of Perretti et al. does indeed lead to biased parameter estimates and poor predictions, the alternative estimation method greatly improves the situation, leading to predictions only little worse than the benchmark of the true parameters, and noticeably better than random on a short time scale (Fig.~\ref{figure: predictive uncertainty}). We did not implement the "model-free" alternatives examined in \cite{Perretti-Modelfreeforecasting-2013}, but in comparing our predictive error with values provided by \cite{Perretti-Modelfreeforecasting-2013}, it is apparent that with our alternative estimation method, the fitted, structurally correct model leads to better short-term predictions than the "model-free" alternatives.

\begin{figure*}
\centering
\includegraphics [width=4.3cm]{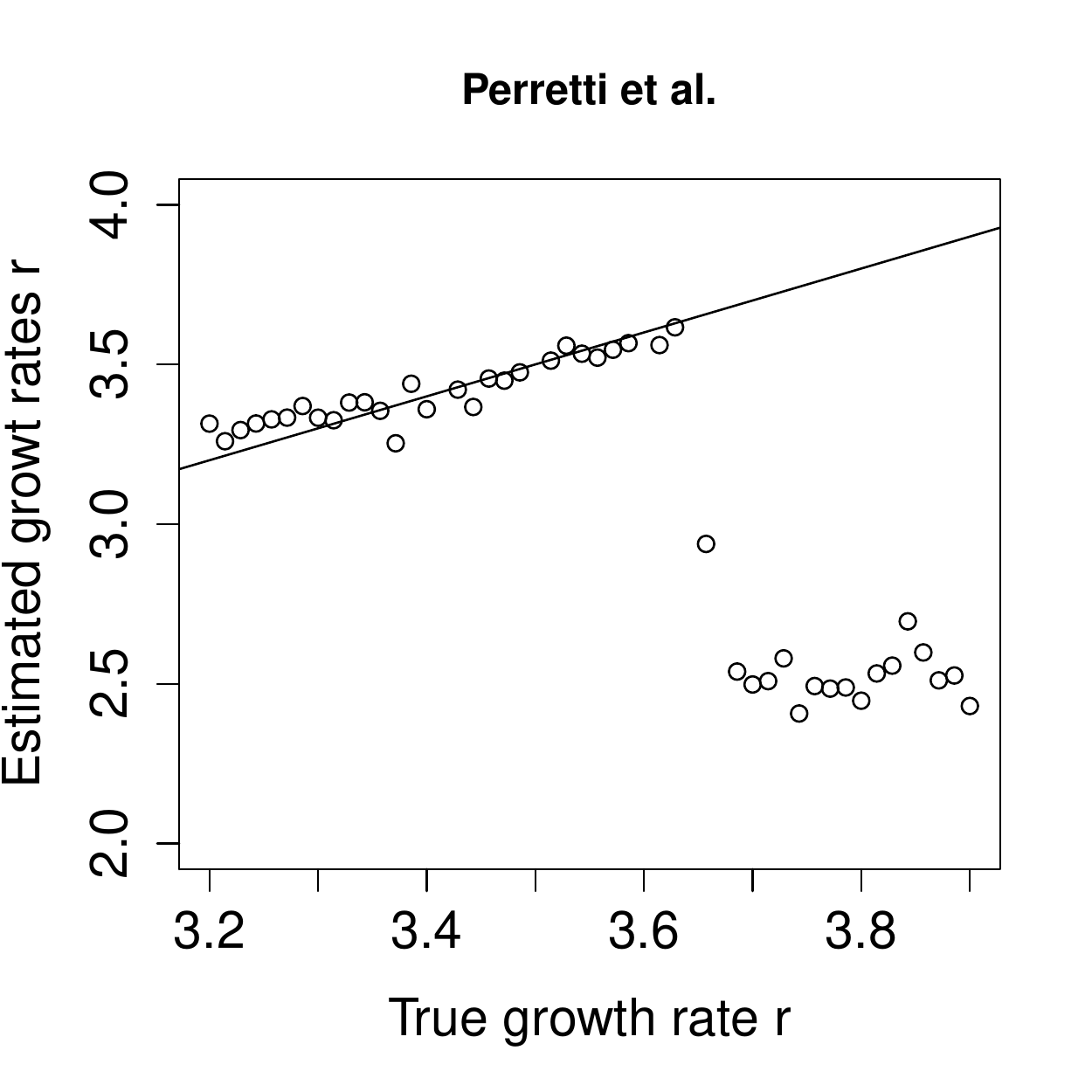} 
\includegraphics [width=4.3cm]{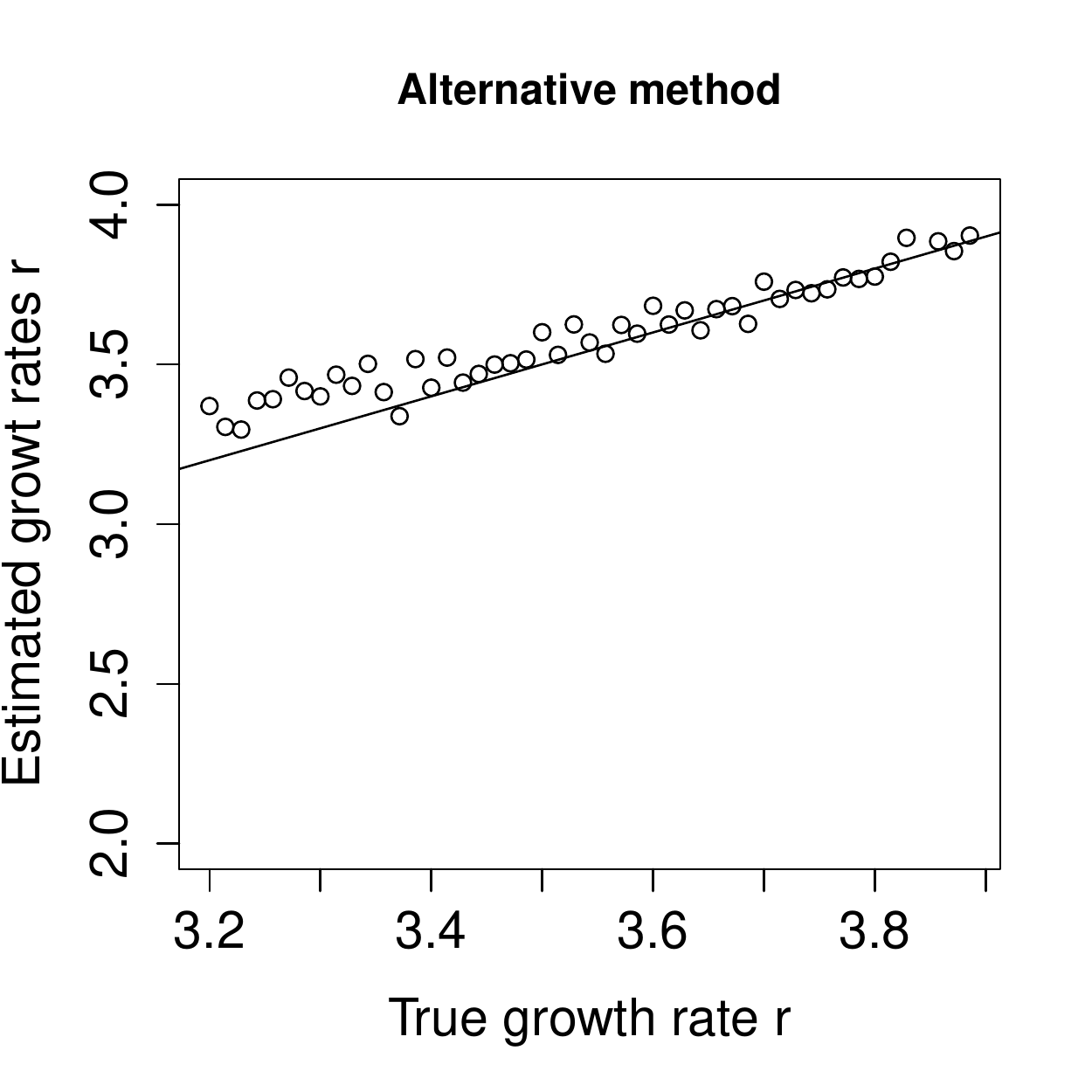} 
\includegraphics [width=4.3cm]{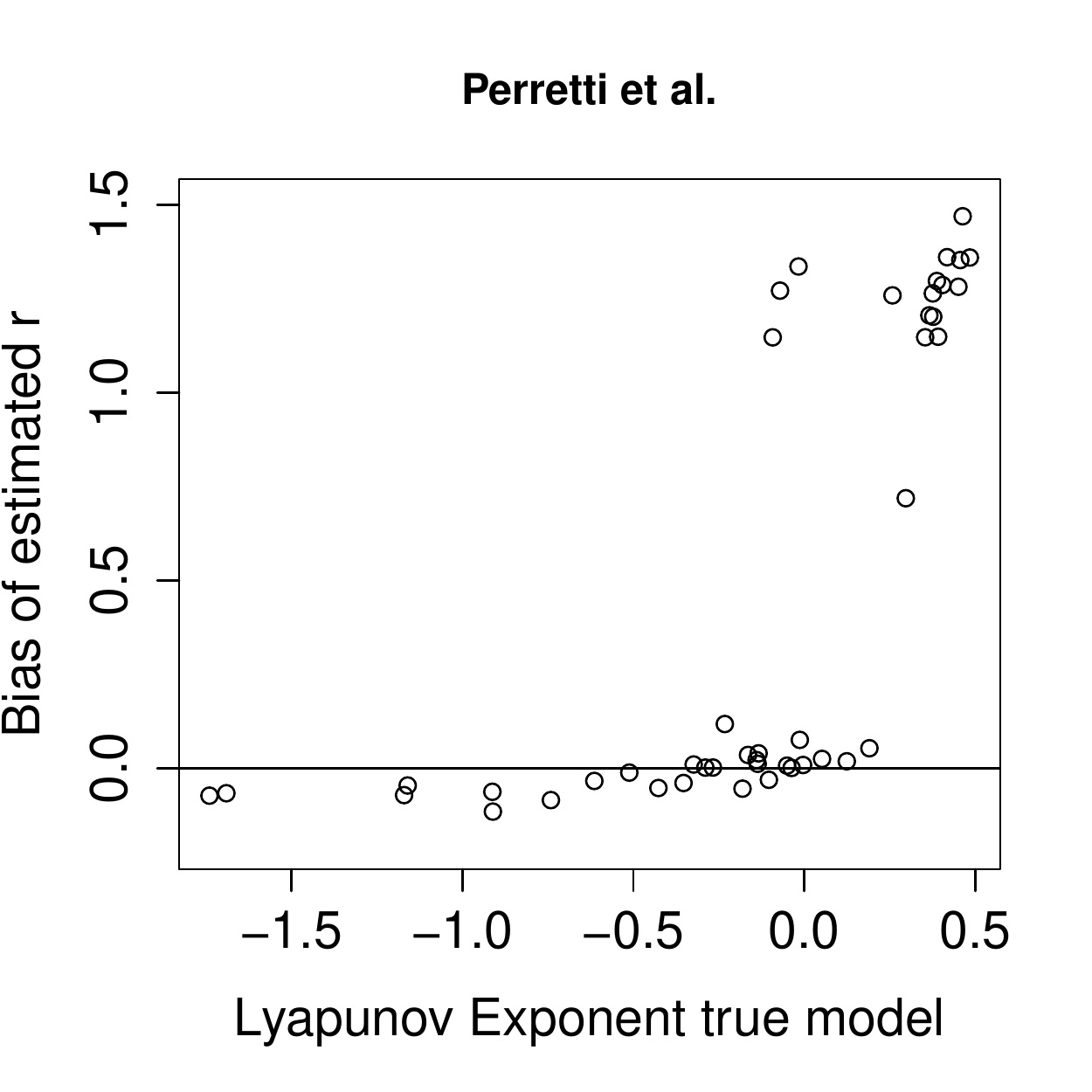} 
\includegraphics [width=4.3cm]{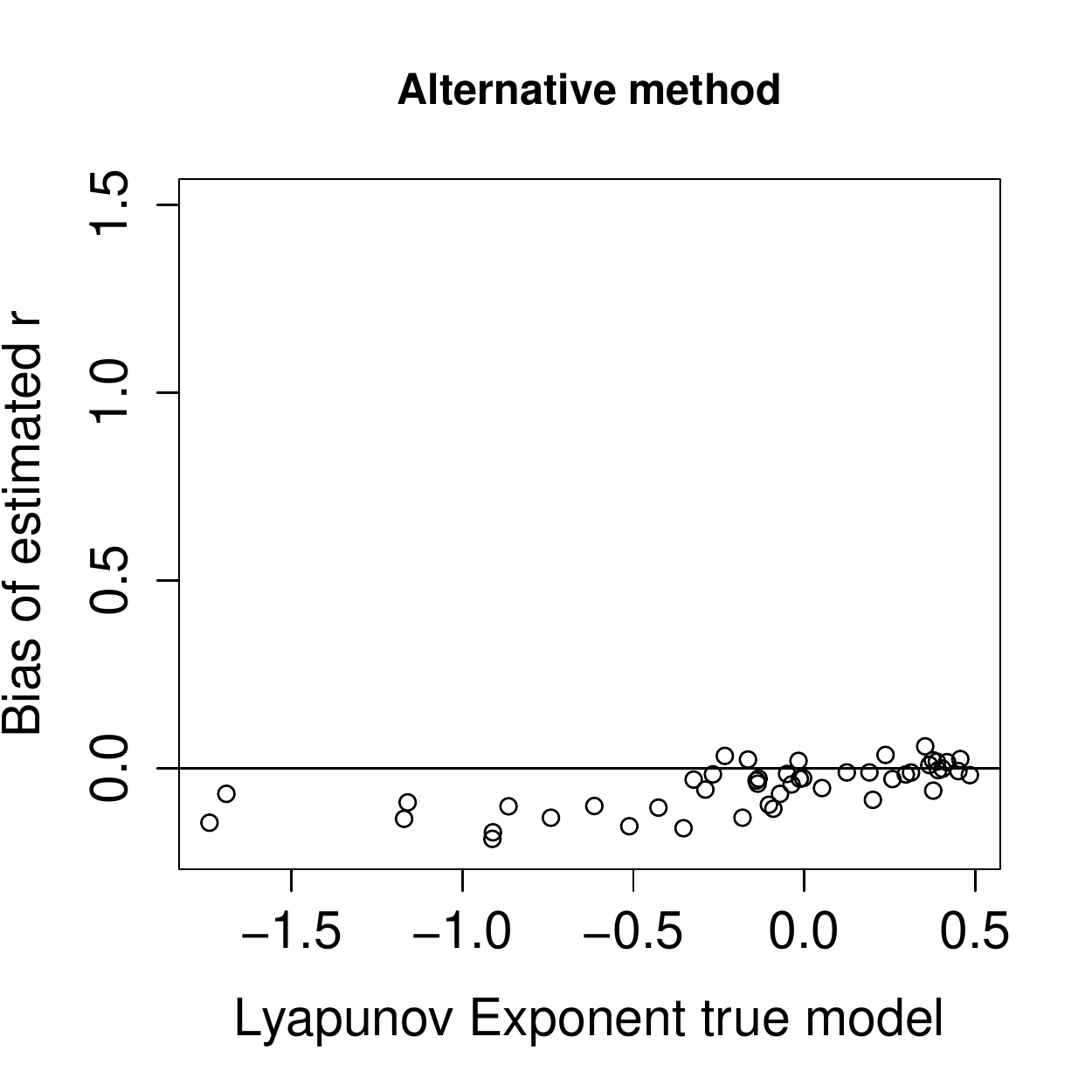} 

\caption{Bias of posterior median estimates for $r$ as a function of the ``true'' $r$, meaning the $r$ fron the model that was used to create the data, and, alternatively, of the Lyapunov exponent of data created with the ``true'' $r$ \label{figure: true vs estimated}} 
\end{figure*}

\subsection{Results: Dependence of the bias on the value of \emph{r}}

Finally, we examined the bias of the parameter estimates across different values of the intrinsic growth rate $r$. We created 50 repetitions with $r$ varying 3.2 and 3.9, and otherwise identical parameters $K=1$, $\sigma^{proc} =0.005$ and $\sigma^{obs} = 0.2$. This choice of parameters allows examining the estimation bias across the transition to the chaotic regime (note that the logistic map shows stable oscillation between $r = 3$ and $r \approx 3.5$, and from that a period-doubling cascade until arriving at chaotic dynamics at $r \approx 3.57$). There is a clear bias in median posterior estimates for the parameter estimation method of Perretti et al. (Fig.~\ref{figure: true vs estimated}) as soon as the model enters the chaotic regime. Parameter estimates are greatly improved when estimates are made with our alternative parameter estimation method. 

\subsection{Conclusions}

In conclusion, our results confirm that the standard Bayesian state-space formulation for estimating the parameters of the logistic map leads to biased parameter estimates. However, a slightly different model specification, where the time series is split up in smaller parts, seems to solve practically all problems - it produces nearly unbiased parameter estimates, fast convergence, is easy to implement, and produces good predictions. Given these results, we see little evidence for the statement that "model-free" forecasting is generally superior to the "true" model for the logistic map. We conjecture that this conclusion can be extended to more complex population models as well.

\section*{Acknowledgements}
We would like to thank J\"orn Pagel and Gita Benadi for helpful comments. All our results were obtained with R version 2.15.2 and JAGS 3.3

\end{document}